\documentstyle[12pt,epsf]{article}
\begin{document}

\begin{flushright}
USM-TH-105
\end{flushright}

\begin{center} \large {\bf Electromagnetic Pulse from Final Gravitational
Stellar Collapse} \\   \end{center}
\normalsize
\vspace*{.15in}
\begin{center}
P.D. Morley \\Development Systems Group B301\\
Veridian Systems Division\\ 14700 Lee Road\\
Chantilly, VA 20151 \\
\vspace*{.15in} Ivan Schmidt\\
Departamento de F\'\i sica, Universidad
T\'ecnica Federico Santa Mar\'\i a, \\
Casilla 110-V,
Valpara\'\i so, Chile\\

\end{center}

\begin{abstract}
We employ an effective gravitational stellar final collapse model
which contains the relevant physics involved in this complex
phenomena: spherical radical infall in the Schwarzschild metric of
the homogeneous core of an advanced star, giant magnetic dipole
moment, magnetohydrodynamic material response and realistic
equations of state (EOS). The electromagnetic pulse is computed both for
medium size cores undergoing hydrodynamic bounce and large size
cores undergoing black hole formation. We clearly show that there
must exist two classes of neutron stars, separated by maximum
allowable masses: those that collapsed as solitary stars
(dynamical mass limit) and those that collapsed in binary systems
allowing mass accretion (static neutron star mass). Our results show that
the electromagnetic pulse spectrum associated with black hole
formation is a universal signature, independent of the nuclear EOS.
Our results also predict that there must exist black holes whose masses
are less than the static neutron star stability limit.

\end{abstract}

\vspace{.25in}

\begin{center} To be published in Astronomy and Astrophysics
\end{center}
\newpage

\section{Introduction}

What is the measurable signal for gravitational collapse? It
certainly produces gravitational waves, but unfortunately these
have shown to be undetectable so far. However, there is another
signal. Since the star has a sizable magnetic field, and since we expect
this field to get quenched very rapidly during gravitational
collapse, a massive electromagnetic pulse (MEMP) is created. We further
show that the power spectrum is unlike any other naturally
occurring phenomena: a wave packet whose wavetrain lasts only
2 milliseconds and whose characteristic power spectrum is a square block.

We present a simulation, within an effective model, of the
iron-core collapse of massive stars which produce type II supernova. Depending
on the core's mass, either a hydrodynamic bounce occurs above nuclear
densities or the collapse races unhindered into the formulation of a stellar
size black hole. Either case produces a MEMP.
Our main contribution in this paper is the calculation of the energy spectrum
of the MEMP. In cases where the collapsing core has mass greater than the dynamical neutron
star mass limit (discussed herein), a hydrodynamic bounce cannot occur and no visible
supernova results. In such cases,
the MEMP is the only electromagnetic signal associated with the gravitational collapse.

One of the unexpected secondary results is the finding that,
in general, there exist two
classes of neutron stars: the first class is made up of solitary neutron
stars which
do not have the opportunity to accrete matter from a binary companion, and
therefore have original masses coming
from the collapse process.
This mass is bounded by the ability of the nuclear equation of state to
produce a hydrodynamic bounce. An unexpected secondary finding is that this imposes
an heretofore unappreciated constraint on candidate nuclear equations of
state. The
maximum mass of the neutron star in this class is the dynamical mass, being
the largest core mass that can undergo a hydrodynamic bounce. We will see
that this dynamical mass limit is considerable less than the static mass limit,
so the bounce does indeed save the core from further collapse due to its
own self-gravity.
The second class of neutron stars are
those that can accrete matter from a binary companion and so their maximum mass
is constrained by the requirement of static stability: the maximum static
neutron star mass. In general, we find that the maximum static mass can
exceed the dynamic mass by as large as 100\%.

A current review of the status of gravitational collapse is given in
Joshi (2000), who discusses spherically symmetric collapse.
In the present paper, we consider spherical radial infall.
As shown in the work of Bocquet et al. (1995), only gigantic magnetic fields
B $> \; 10^{10}$ T = 10 GT (testla = $10^{4}$ gauss, GT = $10^{9}$T)
cause stellar deformation. Since the magnetic fields considered in this paper
fall far below this large critical field, the star is not
deformed by its magnetic field. An interesting followup to the research
presented here, is to apply the effective gravitational collapse model
to stars having non-negligible rotation.

In general, there are three sources for electromagnetic and gravitational radiation associated
with stellar collapse:
the direct radiation phase emitted by the stellar object before the formation
of a black hole, the so-called damped oscillations known as quasi-normal ringing (Iyer 1987)
that are the vibrations of a black hole shaking off its non-zero multipole charge moments, and
the late-time power-law tail (Price 1972;
Cunningham 1978; Leaver 1986; Hod 1999, 2000) underneath the damped oscillations.
Because of the gravitational redshift,
essentially the only energy that survives to infinity comes from the
direct radiation phase, before the black hole forms. In this paper, we calculate the direct
phase electromagnetic energy radiated
by stellar objects that bounce and become stable neutron stars, and
stellar objects so massive that they become black holes.

In order to calculate the electromagnetic radiation produced by the
gravitational collapse, our strategy is the following. During the
collapse the infall kinetic energy decreases while the system does
mechanical work against the internal star pressure. This pressure has, in general, two
components: material (nuclear) pressure and electromagnetic pressure. However, all
published nuclear equations of state have the property that the nuclear pressure
is orders of magnitude larger than the electromagnetic pressure; indeed, only for
magnetic fields greater than $10^{18}$ Gauss does the electromagnetic pressure approach parity
with the nuclear pressure. Using conservation of energy, the continuity equation and the
material equation
of state, we can calculate how the star radius changes with time.
If we further assume that the stellar magnetic field can be
approximated by a dipole field, and that the lines of magnetic
field are frozen in the material and carried along with it, we can
relate the change in the star's radius with time to the change in
its effective magnetic moment. The usual dipole radiation formula
then gives the energy spectrum of the radiation.

\section{The Effective Gravitational Collapse Model}

\begin{figure}[htb]
\begin{center}
\leavevmode {\epsfysize=5.5cm \epsffile{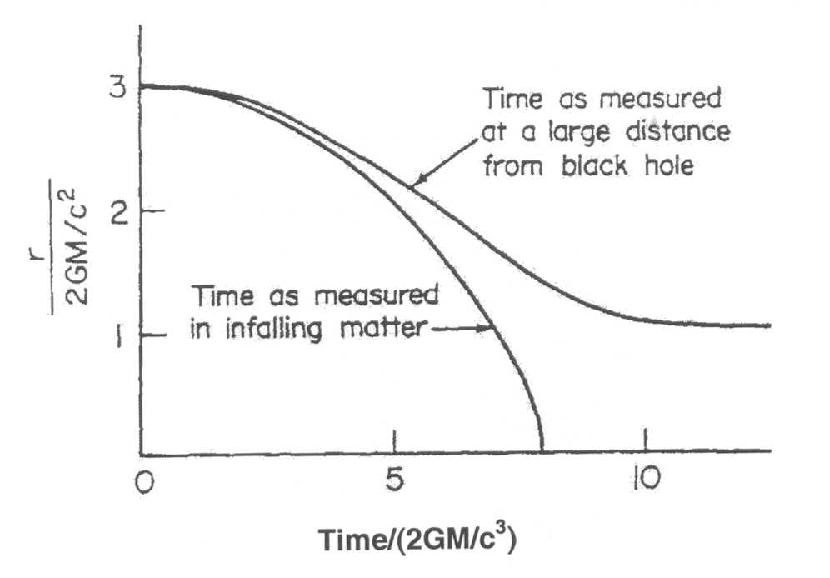}}
\end{center}
\caption[*]{\baselineskip 13pt Difference between proper and
Schwarzschild times. } \label{1471f1}
\end{figure}

We simulate the collapse to the hydrodynamic bounce, or to twice
the gravitational radius, $a_{g} = 2GM/c^2$, if no bounce occurs. In the former
case,
the bounce cuts off the MEMP, while in the latter case, final
black hole formation at radius = $a_{g}$ occurs almost
instantaneously later after the simulation ends. Because of the gravitational
redshift, the final stellar trajectory from 2$a_{g}$ to $a_{g}$ adds little to the
observable energy at infinity.
Thus the MEMP
calculated is expected to be an accurate profile. The reason why we stop at 2$a_{g}$
is because, while the Schwarzschild metric is the exterior
metric during gravitational collapse and as such, is the clock time in the
effective collapse model, the radiation process
must be calculated in proper time.
In Fig.\ 1, we plot the
difference between proper time and Schwarzschild time. Only when the
core radius becomes less than 3$a_{g}$ is there a difference in the two clocks.
By stopping
the simulation at 2$a_{g}$, we can still reliably calculate the core surface
as a function of proper time, by using the Taylor expansion employing
the Schwarzschild time solution of the collapse model. Indeed, one can calculate
from this figure that the time error is 8.364 microseconds
$\times$ X, where ${\cal M}/{\cal M}_{\odot}$ = X, with $\cal M$
being the core mass. At the same time, we fold
in the gravitational and Doppler redshift, since the difference between the
Schwarzschild time
and proper time leads to a gravitational redshift of the produced
radiation.

The collapse process from the initial starting configuration of
the iron-core mass density ($\rho$) of $\rho = 1 \times 10^{12}$
gm/cm$^{3}$ and inward speed $v/c \sim 2/3 \times 10^{-2}$ to
either hydrodynamic bounce or twice $a_{g}$ takes
less than 3 milliseconds. In contrast, the characteristic time
scale for neutrino diffusion out of the core is $\sim 2\rightarrow
10$ seconds. This $10^{3}$ times larger neutrino diffusion scale
means that these particles cannot influence the bounce or the MEMP.
The role of the
neutrinos is not to carry away the binding energy of the collapsing
matter, as it sometimes is erroneously stated.
Rather, the collapsing matter must do work against the equation of
state and this energy is stored as {\it internal energy}, entirely
analogous to a weight compressing a spring. This stored energy is
not lost, but remains primarily as elevated Fermi levels of electrons and nucleons.
Thus the energy the
neutrinos actually carry away is the energy from nuclear beta
decay, relaxed elastic energy and thermal heat energy. The
difference in the time scales for neutrino diffusion and
hydrodynamic infall means that the iron-core collapse is
adiabatic, allowing a concise formulation for the conservation of
energy. Previous work, Morley (1999), showed that neutrino
transport in the iron-core was due to elastic scattering
and typically $\sim$ 1000 collisions occurred before escape. Our
collapse code in the present paper shows beyond a doubt, that if
no hydrodynamic bounce occurs, these diffusing neutrinos are
trapped in the infalling nuclear material and are carried into the
resulting black hole.
Thus observation of neutrinos associated
with particular supernova proves that the initial infall did not
form a black hole.

Another physical insight into the problem is the recognition that the iron-cores
have flat density profiles. In Fig.\ 2, we show the neutron star density
profile using the Friedman-Pandharipande equation of state, Straumann (1992).
Thus the approximation that the star collapses
homogeneously is excellent ($t$ = Schwarzschild time):
$\rho(\vec{r},t) =\rho(t)$. In this paper,
we treat spherical radial infall. As mentioned earlier, a very
interesting follow-on to this research is to remove this constraint and
allow core rotation in the effective collapse model.

\begin{figure}[htb]
\begin{center}
\leavevmode {\epsfysize=5.5cm \epsffile{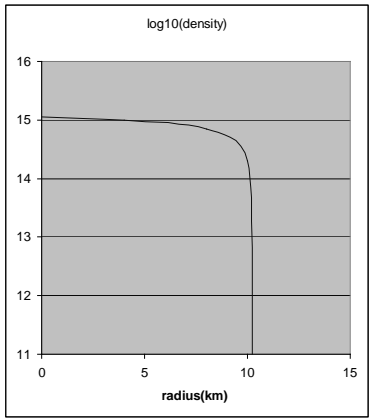}}
\end{center}
\caption[*]{\baselineskip 13pt Neutron star density profile using
the Friedman-Pandharipande equation of state, Straumann (1992). }
\label{1471f2}
\end{figure}

Further simplifications can be achieved by using the fact that the electrical
conductivity of the iron-core is very high, nearly infinite.
This allows the magnetohydrodynamic
approximation: the lines of magnetic force are frozen into the material and
are carried along with it. In this paper, we approximate the iron-core
stellar
magnetic field by a magnetic dipole of moment $\vec{m}$. We use the model of
a uniformly
magnetized sphere of magnetization per unit volume $\vec{M}$. The relation of
$\vec{m}$ to $\vec{M}$ is
    \begin{equation} \vec{m} = \frac{4 \pi}{3} a^3 \vec{M}
    \end{equation}
where $a$ is the stellar radius. $\vec{B}_{out} = \vec{H}$ outside the sphere and
is a magnetic dipole field:
   \begin{equation} \vec{B}_{out} = \frac{4 \pi}{3} a^3 |\vec{M}|
   \{ \frac{2 \cos \theta}{r^{3}}\hat{r} + \frac{\sin \theta}{r^{3}} \hat{\theta}
   \}
   \end{equation}
while inside the core
   \begin{equation} \vec{B}_{in} = \frac{8 \pi}{3} \vec{M} \; .
   \end{equation}

\begin{figure}[htb]
\begin{center}
\leavevmode {\epsfysize=5.5cm \epsffile{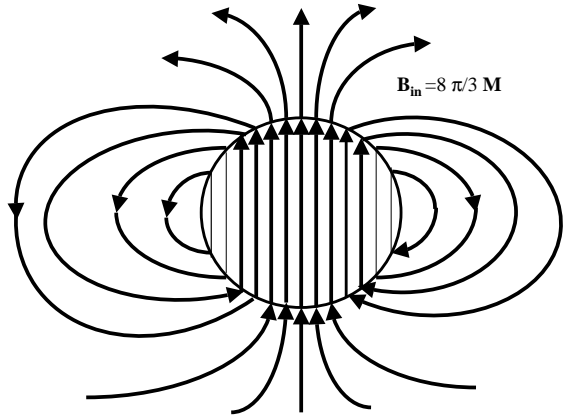}}
\end{center}
\caption[*]{\baselineskip 13pt Stellar magnetic dipole field. }
\label{1471f3}
\end{figure}

This is shown in Fig.\ 3. In general, $a = a(t)$ and
$\vec{M}=\vec{M}(t)$. The magnetohydrodynamic material response is
the constancy of the magnetic flux through the polar region, Fig.\
4

  \begin{equation} \pi  a^{2}(t_{1})\sin^{2} \theta B_{in}(t_{1}) = \pi
  a^{2}(t_{2}) \sin^{2} \theta B_{in}(t_{2}), \; |\vec{B}_{in}| = {B}_{in}
  \end{equation}

giving
  \begin{equation}  a^{2}(t_{1})|\vec{M}(t_{1})| =
  a^{2}(t_{2})|\vec{M}(t_{2})| = {\rm constant} = C_{m} \; . \end{equation}

\begin{figure}[htb]
\begin{center}
\leavevmode {\epsfysize=5.5cm \epsffile{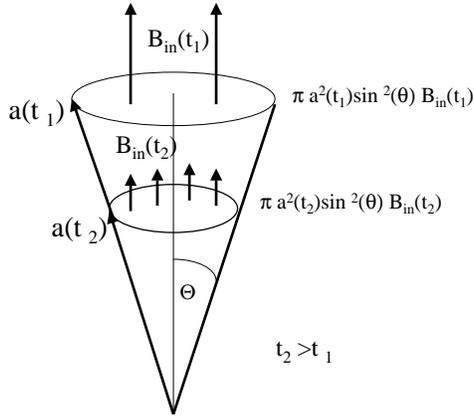}}
\end{center}
\caption[*]{\baselineskip 13pt Constancy of the magnetic flux
through the polar region. } \label{1471f4}
\end{figure}

Therefore
  \begin{equation} \vec{m} = \frac{4 \pi}{3}aC_{m} \hat{z}
  \end{equation}  and
  \begin{equation} \ddot{m}(t) = \frac{4 \pi}{3}C_{m} \ddot{a}(t),  \;
  m = |\vec{m}| \; .
  \end{equation}
$C_{m}$ is fixed by astrophysical data. Young pulsars (e.g. PSR B0154+61)
have the polar field strength $B \cong 2$GT. Since this is over a flux
area of $10 \times 10$ km$^{2}$, the constant is $C_{m} \simeq 8 \times 10^{25}$
cm$^{2} \cdot$Gauss.

Thus we have obtained a relation between the second order time derivative of the
stellar effective
magnetic moment and the star's surface acceleration, which will allow us to
obtain the
power radiated once we know the acceleration. This will be
obtained by a numerical simulation, whose ingredients we explain
in the next sections.

\subsection{Conservation of Matter}
The conservation of baryon number is given by the equation
   \begin{equation} \frac{dn_{bar}}{d \tau} = -n \bigtriangledown \cdot u
   \end{equation}
where $n_{bar}$ is the baryon density, $\tau$ is proper time,
$\bigtriangledown$
is the 4-derivative and $u$ is the 4-velocity. As long as $\tau \simeq$ t and
extreme relativistic speeds are not encountered, then the conservation of
baryon number reduces to the
the conservation of matter:
  \begin{equation} \vec{\nabla} \cdot (\rho \vec{v}) + \frac{\partial \rho}
  {\partial t} = 0 \; .
  \end{equation}
The only region in the collapse phenomena where the two conservation laws
begin to
differ is near twice $a_{g}$.
Though we treat $\rho$ as uniform throughout the spherical volume, it is changing
in time $t$ so $ \frac{\partial \rho}{\partial t} = \frac{d \rho}{dt} \neq 0$.
Thus $\vec{v}$ cannot be a
constant. With
$\vec{\nabla} \cdot (\rho \vec{v}) = \rho \vec{\nabla} \cdot \vec{v} =
\frac{\rho}{r^{2}} \frac{d}{dr}(r^{2}v_{r}) $
where $\vec{v} = v_{r}\hat{r}$, the conservation of matter equation becomes
  \begin{equation} \frac{d \rho}{dt} +\frac{\rho}{r^{2}}
  \frac{d}{dr}(r^{2}v_{r})=0
  \; .  \end{equation}
This has the physical solution
   \begin{equation} v_{r} = -\frac{r}{3 \rho} \frac{d \rho}{dt} \label{con} \; .
   \end{equation}

\subsection{Conservation of Energy}
The energy $E_{if}$ available to do work if the matter collapses from mass
density $\rho_{i}$ to $\rho_{f}$ is
  \begin{equation} E_{if} = PE_{i} + KE_{i} - PE_{f}
  \end{equation}
where $KE_{i}$, $KE_{f}$ is the kinetic energy at the initial, final density
respectively, and $PE_{i}$, $PE_{f}$ is the gravitational self-energy of
the core with initial, final mass density respectively. The work $(\cal W)\; > \; 0$
done against
the equation of state in compressing the matter is
  \begin{equation} {\cal W} = -\int_{V_{i}}^{V_{f}} P dV, \; P = {\rm pressure}.
  \end{equation}
Here $V_{i}, \; V_{f}$ are the initial and final core volumes respectively.
The core will bounce when or if it reaches a density such that
  \begin{equation} E_{if} = {\cal W} \label{bou} \; .
  \end{equation}
For realistic equations of state, we expect a bounce for core masses of the
order of a solar mass.

\subsection{Gravitational Self-Energy}
A core of radius $a$ in the Schwarzschild metric has gravitational potential
energy, Weinberg(1972)
  \begin{equation} PE = \int_{0}^{a} 4 \pi r^{2} \{ 1 - \frac{1}
  {\sqrt{1-\frac{2G{\cal M}(r)}{c^{2}r}}} \} \rho_{e} dr
  \end{equation}
where $c$ is speed of light, $\rho_{e}$ is the energy density of matter and
${\cal M}(r)$
is the amount of mass within radius $r$.
Since ${\cal M}(r) = \frac{r^{3}}{a^{3}}{\cal M}$
and $\rho_{e} = \frac{3{\cal M}c^{2}}{4 \pi a^{3}}$, this reduces to
  \begin{equation} PE = \rho_{e} \int_{0}^{a} 4 \pi r^{2}
  \{1 - \frac{1}{\sqrt{1-\frac{2G{\cal M}r^{2}}{c^{2}a^{3}}}}  \} dr
  \end{equation}
which can be integrated to give
  \begin{equation} PE = {\cal M}c^{2} +\frac{2 \pi \rho_{e} a}{\kappa} \{
  \sqrt{1-\kappa a^{2}} -\frac{1}{a \sqrt{\kappa}} \arcsin{a\sqrt{\kappa}} \label{grv}
  \end{equation}
where $\kappa = \frac{2G{\cal M}}{c^{2}a^{3}}$, $G$ = gravitational constant.

\subsection{Kinetic Energy of Infall}
The equation of state (EOS) of matter (to be discussed in more detail later) can
describe ``hot" or ``cold" nuclear matter. In general, the thermal energies
that are reached in the collapse phenomena are a small fraction of the
gravitational binding energy experienced by the particles. If the pressure $P$
in Eq.\ (13) describes a ``hot" EOS, then some of the kinetic energy + potential energy
of infall is turned into heat. The numerical procedure (to be discussed in
detail in a later section) takes small time slices ($\Delta t$) and computes the new
kinetic energy after work against the EOS is determined. The equation is
  \begin{equation} KE(t+\Delta t) = KE(t) +PE(t) -PE(t+\Delta t)
  +\int_{V(t)}^{V(t+\Delta t)} PdV
  \label{kin} \; .
  \end{equation}

Thus the effective gravitational model takes into account any
desired heat generation through the inputed EOS. We need the
instantaneous kinetic energy of a collapsing iron-core in the
Schwarzschild metric. We
want the expression to be accurate to O($v_{r}^{4}/c^{4}$).

To order O($v_{r}^{2}/c^{2}$), the instantaneous kinetic energy is
   \begin{equation} KE = \int_{0}^{a} 4 \pi r^{2}(\frac{\rho_{e}(r)}{2})
   \frac{v^{2}_{r}(r)}
   {c^{2}}{1 \over \sqrt{1 - \frac{2G{\cal M}(r)}{c^{2}r}}} dr
   \end{equation}
where
   \begin{equation} v^{2}_{r} (r) = r^{2}(\frac{1}{3 \rho}\frac{d \rho}{dt})^{2}
   =r^{2}{\cal B}, \; {\cal B} = (\frac{1}{3 \rho}\frac{d \rho}{dt})^{2} \; .
   \end{equation}
This becomes
   \begin{equation} KE = 2 \pi \rho_{e} \frac{{\cal B}}{c^{2}} \int_{0}^{a} \frac{r^{4}dr}
   {\sqrt{1 - \kappa r^{2}}} \; ,
   \end{equation}
Computing this integral:
  \begin{equation} KE =  2 \pi \rho_{e} \frac{{\cal B}}{c^{2}} \{ -\frac{a^{3}}{4
  \kappa}
  \sqrt{1 - \kappa a^{2}} -\frac{3}{8 \kappa^{2}} a \sqrt{1 - \kappa a^{2}}
  +\frac{3}{8 \kappa^{5/2}} \arcsin a\sqrt{\kappa} \} \; .
  \end{equation}
The kinetic energy to O($v_{r}^{4}/c^{4}$) is
   \begin{equation} KE = \int_{0}^{a} 4 \pi r^{2}(\frac{\rho_{e}(r)}{2}) [
   \frac{v^{2}_{r}(r)}{c^{2}} -\frac{1}{8}\frac{v^{4}_{r}(r)}{c^{4}} ]
   {1 \over \sqrt{1 - \frac{2G{\cal M}(r)}{c^{2}r}}} dr \; .
   \end{equation}
This extra piece can be integrated in the same manner as the first piece,
and using the parameters $\dot{a}$ (speed of surface) and
$\lambda = \kappa a^{2}$ we find that the kinetic energy to
O($v_{r}^{4}/c^{4}$) is
    \begin{eqnarray} KE & =&  {\cal M}c^{2} \{ \frac{3}{2}(\frac{\dot{a}}{c})^{2}
    \frac{1}{\lambda^{5/2}} [ -\frac{1}{4} \lambda^{3/2} \sqrt{1-\lambda}
    -\frac{3}{8}\sqrt{\lambda}\sqrt{1-\lambda}+\frac{3}{8}
     \arcsin \sqrt{\lambda}]  \nonumber \\
     & &
     +\frac{15}{2} (\frac{\dot{a}}{c})^{4}
     \frac{1}{\lambda^{7/2}} [ \sqrt{1-\lambda}( \frac{\sqrt{\lambda}}{64}
     +\frac{\lambda^{3/2}}{96} +\frac{\lambda^{5/2}}{120}) -\frac{1}{64}
     \arcsin \sqrt{\lambda}] \label{kin2}
     \end{eqnarray}

\subsection{Nuclear Equation of State (EOS)}
The hydrodynamic bounce is independent of the equation of state
below nuclear density (nuclear density $\simeq 2 \times 10^{14}$
gm/cm$^{3}$). We know this because if we put in zero pressure in
Eq.\ (13) below nuclear density, the bounce properties hardly
change at all.
The hot thermal gas
kinetic EOS below nuclear density is not responsible for the
hydrodynamic bounce, and this can be shown by calculating the work ${\cal
W}$.
In this paper, we use two completely different nuclear EOS, and while
the contribution of the pressure below nuclear densities is
negligible, we still hook them to EOS of degenerate electronic
matter used frequently in collapse calculations, Baym (1971).

The different regimes for the nuclear EOS are listed in Table 1.
   \begin{table}[h]
   \begin{tabular}{||c|c|c||} \hline
   Class & Property & Applicability \\ \hline
   \mbox{} & \mbox{} & \mbox{} \\
   normal matter & N $\simeq$ Z & atomic nuclei \\
   iron-core & N $\simeq$ 2Z & proto-neutron star \\
   neutronic & N $>>$ Z  & cold neutron star \\ \hline
   \end{tabular}
   \caption{The three regimes for the nuclear EOS. The first and third are
   in beta decay equilibrium, but the middle is out of equilibrium.
   Here N = the number of neutrons and Z = the number of protons.}
   \end{table}

In this paper, two different nuclear EOS were used in the collapse
calculations: the ``Argonne AV14 + UVII", Wiringa (1995) and the QCD,
Kislinger (1978).

Perturbation theory in quantum chromodynamics (QCD) breaks down at
$\sim$ twice nuclear density, where it must be joined to phenomenological
EOS. Both EOS used here are joined to the Baym EOS, as mentioned earlier. At
some time in the future, the nuclear EOS will be known accurately, and then
collapse calculations will lead to predictions with less error.

\begin{figure}[htb]
\begin{center}
\leavevmode {\epsfysize=5.5cm \epsffile{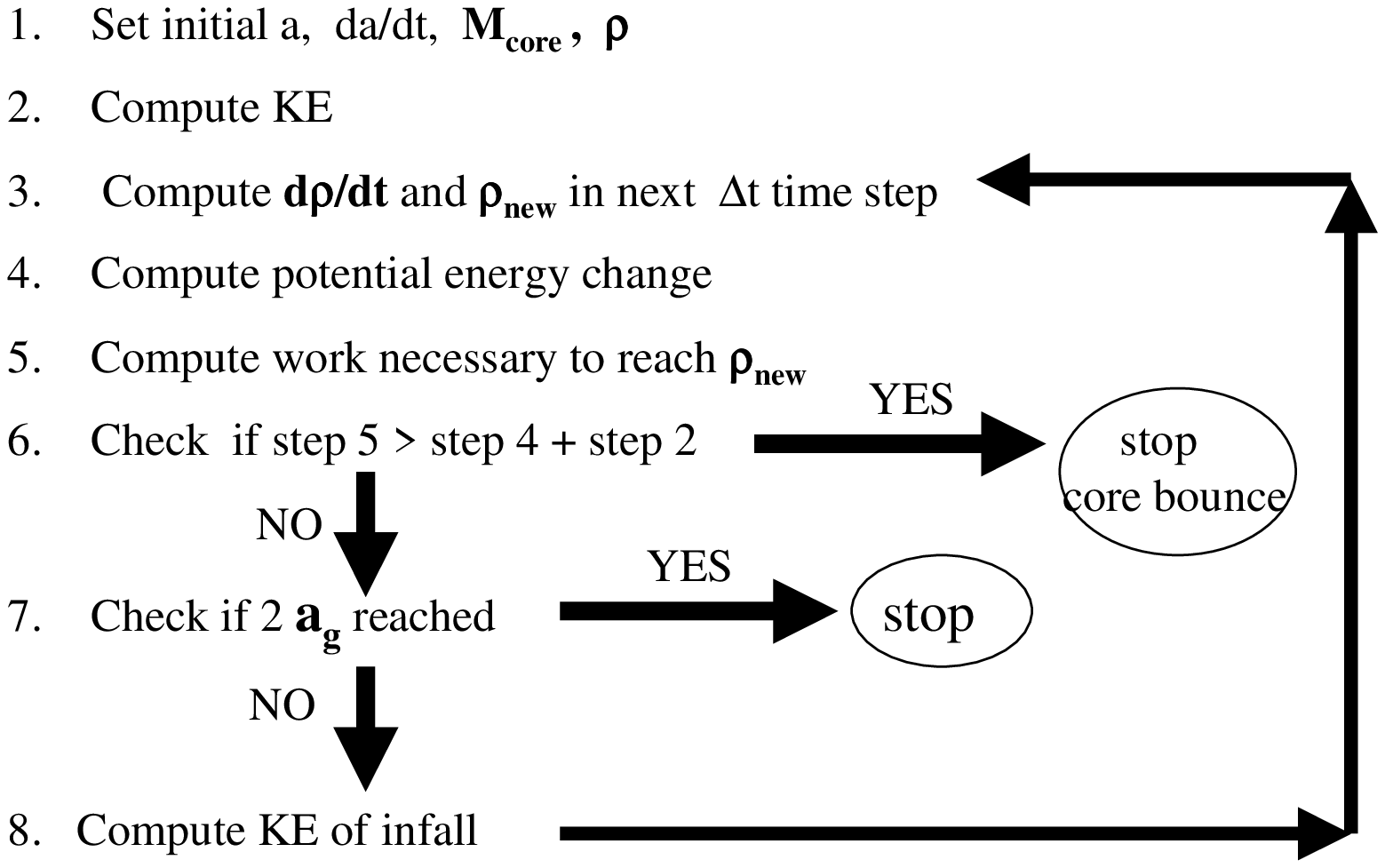}}
\end{center}
\caption[*]{\baselineskip 13pt Code logic. } \label{1471f5}
\end{figure}

Now we have all the ingredients for the numerical simulation. Once
the initial stellar radius, surface velocity, mass and density are
given, we can compute $d\rho/dt$ using the continuity equation
(\ref{con}), and therefore the new density and potential energy
(equation \ref{grv}) in the next $\Delta t$ step.  Then we
calculate the work necessary to reach this new density using the
corresponding EOS, and check whether the bounce condition given in
equation (\ref{bou}) is met. If it is, which means that the new
kinetic energy vanishes, there is a bounce and we stop the
simulation. If it is not, we continue by checking also whether we
have reached twice $a_{g}$, in which case we also stop. If we have
not reached that radius we compute the new kinetic energy
(\ref{kin}), which allows us in turn to get the new surface
velocity solving equation (\ref{kin2}) and then use the continuity
equation again, and so on. This logic is described in Fig.
\ref{1471f5}. Notice that the desired quantity, the surface
acceleration, is obtained directly from the continuity equation
(\ref{con}), as

\begin{equation}
\ddot a = \frac{4}{9}\frac{a}{{\rho ^2 }}\left( {\frac{{d
\rho }}{{dt}}} \right)^2  - \frac{a}{{3\rho }}\frac{{d^2
\rho }}{{dt^2 }}
\end{equation}

\section{Radiation}
The MEMP calculated here is magnetic dipole radiation; it is not the emission
of a black-body at some temperature and radius. The MEMP, while it exists,
completely swamps thermal
emission. Our goal is to calculate the MEMP energy spectrum.
The energy radiated in all directions per unit time, in Gaussian units,
is, Landau(1975):
   \begin{equation} \frac{dE}{d\tau} = \frac{2}{3 c^{3}}|\ddot{m}(\tau)|^{2} \; .
   \end{equation}
We introduce the Fourier transforms $m(\omega)$
    \begin{equation} m(\tau) = \frac{1}{\sqrt{2 \pi}} \int m(\omega)
    e^{i \omega \tau} d \omega
    \end{equation}

    \begin{equation} m(\omega) = \frac{1}{\sqrt{2 \pi}} \int m(\tau)
     e^{-i \omega \tau}  d\tau \; .
     \end{equation}
Then
  \begin{equation} \ddot{m}(\tau)= -\frac{1}{\sqrt{2 \pi}} \int \omega^{2}
  m(\omega) e^{i \omega \tau} d \omega \; .
  \end{equation}
Taking the integration over time from $-\infty$ to $+\infty$ gives the total
energy radiated
  \begin{equation} E = \frac{2}{3 c^{3}} \int \omega^{4} |m(\omega)|^{2}
  d \omega \end{equation}
and the energy spectrum
  \begin{equation} \frac{dE}{d \omega} = \frac{2}{3 c^{3}}
  \omega^{4} |m(\omega)|^{2} \; .
  \end{equation}
\subsection{Roll-on-Roll-off (RORO) Function}
The core's dipole moment is
   \begin{equation} m(\tau) = \frac{4 \pi}{3} C_{m} a(\tau)
   \end{equation}
where $a(\tau)$ is the core's radius in km at proper time $\tau$. As discussed
in the later sections, we fit $a(\tau)$
to a polynomial curve and compute
$\ddot{m}(\tau)$. Then we integrate analytically to obtain the total
energy release. To know the energy spectrum, though, and to include the Doppler and gravitational
redshifts, we must compute
the Fourier transform. If we put in Heaviside functions in time
for the beginning and ending of the infall, the Fourier transform
will not give rise to a finite energy spectrum because the
derivative of a step function is a delta function. Physically we
must impose the following boundary conditions: from time $-\infty$
we had the initial radius and at time $+\infty$ we have the final
radius, and smooth transition between them in the infall period.
The analogous problem occurs in particle scattering. Thus we need
a Roll-on-Roll-off (RORO) function to accomplish this mathematically. Any
suitable RORO function will do and gives the same
answer. Thus the time dependent $m(\tau)$ becomes (with $C_{0} =
\frac{4 \pi}{3} C_{m}$)

   \begin{eqnarray}  m(\tau) &  \Rightarrow & C_{0}\frac{1}{2}a_{max}
   (1-\tanh ( \frac{\tau}{ \mu {\rm s}} ))   \nonumber \\
   & & +.25(1+\tanh ( \frac{\tau}{ \mu {\rm s}}))
   \times (1-\tanh( \frac{\tau-T}{ \mu {\rm s}}))m(\tau) \nonumber \\
   & & + C_{0}\frac{1}{2}a_{min}(1+\tanh ( \frac{\tau-T}{ \mu {\rm s}})) \; .
   \end{eqnarray}

In the above equation, $m(\tau)$ on the right-hand-side is the fitted polynomial
function in proper time $\tau$ (fifth order polynomial - fitted from the collapse code
output of the acceleration), $\tau$ and $T$ = collapse time, are in microseconds, and $a_{max}$
and $a_{min}$ are respectively the starting and ending core radius in km.
Any derivation of the
gravitational or electromagnetic power spectrum must use a RORO function. As we
will see in the following sections, the RORO function is also critical to
deriving the redshifts.

\subsection{Total Energy}
We fit the acceleration collapse solution to a polynomial function
    \begin{equation} a(\tau) = {\rm km} \sum_{i = 0}^{5} b(i)(\tau/
    \mu {\rm s})^{i}
    \end{equation}
where $\tau$ is in microseconds. Then
  \begin{equation} m(\tau) = C_{0} {\rm km}
  \sum_{i = 0}^{5} b(i)(\tau/  \mu {\rm s})^{i}
  \end{equation}
  and
  \begin{equation} |\ddot{m}(\tau)|^{2} = C_{0}^{2} {\rm km}^{2} (\frac{1}
  { \mu {\rm s}})
  ^{4} \sum_{i=2}^{5} b(i) (\frac{\tau}{\mu {\rm s}})^{i-2}
  \sum_{j=2}^{5} b(j)
  (\frac{\tau}{\mu {\rm s}})^{j-2}   \; .
    \end{equation}
The total energy emitted in the direct phase of the collapse is then
  \begin{equation} E = \frac{2 C_{0}^{2} {\rm km}^{2}}{3 c^{3}}(\frac{1}{\mu {\rm s}})
  ^{3} \sum_{i=2}^{5} \sum_{j=2}^{5} b(i)b(j) (\frac{T}{\mu {\rm s}})^
  {i+j-3}\frac{1}{i+j-3} \; .
  \end{equation}
The $b(i)$ are fitted by a least-squares program. $T$ is the collapse time in microseconds.

When the collapse extends below 3 $a_{g}$, we need to use multiple RORO
functions to separate out the contribution of each surface to the radiation
spectrum, since each surface has a different gravitational redshift
and Doppler shift. Then the energy radiated from each surface radius
can be corrected by the correct attenuation factor, as perceived from
an observer far away.
\subsection{Fourier Transform}

\begin{figure}[htb]
\begin{center}
\leavevmode {\epsfysize=5.5cm \epsffile{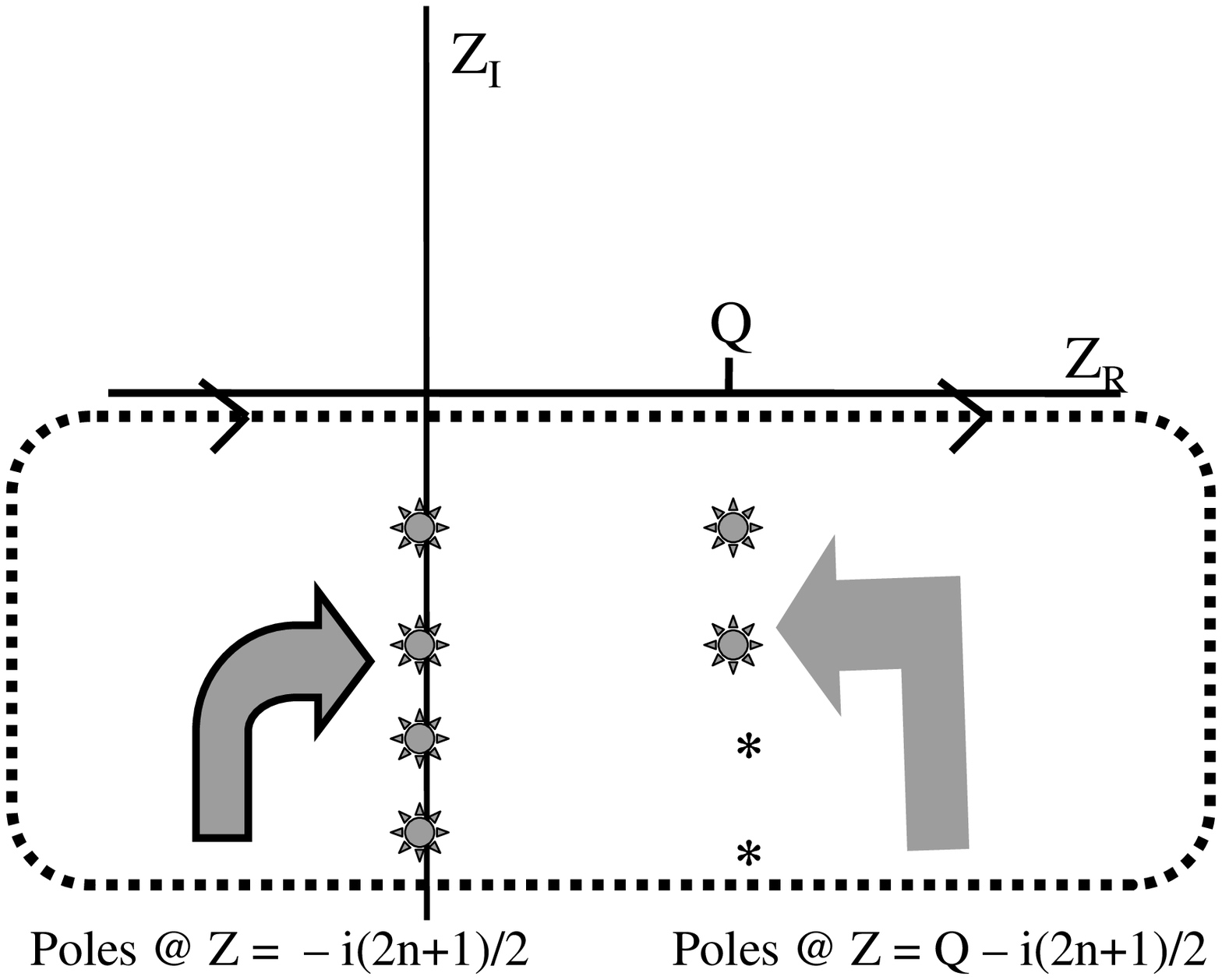}}
\end{center}
\caption[*]{\baselineskip 13pt Integration contour. }
\label{1471f6}
\end{figure}

We must compute $m(\omega)$ where $m(\omega)$ is given by Eq.\ (28).
Letting $s \equiv \omega \cdot \mu {\rm s}$, $x \equiv \tau/ \mu {\rm s}$ and
$Q \equiv T / \mu {\rm s}$, we have
     \begin{eqnarray} m(\omega) &  = &  \mu {\rm s \; km} \; C_{0} \frac{1}{\sqrt{2 \pi}}
     \int_{-\infty}^{+\infty} dx \; e^{-isx} \{ .5\frac{a_{max}}{ {\rm km}}
     (1-\tanh x) \nonumber \\
     & & + .25(1+\tanh x)(1-\tanh(x-Q))\sum_{i=0}^{5}
     b(i)x^{i} \nonumber \\
     & & +.5\frac{a_{min}}{{\rm km}}(1+\tanh(x-Q)) \}
     \; .
     \end{eqnarray}

This integral can be done using contour integration in the complex energy plane.
Letting
$x = Z_{r} + iZ_{i}$ (complex), we have damping for $Z_{i}<0$. Poles in the
lower half of the complex plane occur at
$Z_{r}=0,\;  Z_{i} = -\frac{2N+1}{2} \pi$
and at $Z_{r}=Q,\;  Z_{i} = -\frac{2N+1}{2} \pi$. We integrate
clockwise along the contour of Fig.\ 6. to obtain the answer:
   \begin{eqnarray} m(\omega) & = & \frac{\mu {\rm s \;  km} \; C_{0}}
   {\sqrt{2 \pi}} (-2 \pi i) \{ \sum_{N=0}^{\infty} e^{-s\pi \frac{2N+1}{2}}
   [ -.5(a_{max}/{\rm km})  \nonumber \\
   & & +.25 \frac{1 + \tanh Q}{\tanh Q}\sum_{j=0}^{5}b(j)
   (-i \pi (\frac{2N+1}{2}))^{j}] \nonumber \\
   & & + \sum_{N=0}^{\infty}e^{-isQ -s \pi(\frac{2N+1}{2})} [-.25
   \frac{1 + \tanh Q}{\tanh Q}
   \cdot \sum_{j=0}^{5}b(j) \cdot  \nonumber \\
    & &  (Q-i \pi (\frac{2N+1}{2}))^{j}
   +.5(a_{min}/{\rm km})] \}
   \end{eqnarray}
The presence of the exponentials signals that the integration over $\omega$
in Eq. (30) is indeed finite.

\subsection{Redshift}
The difference between the proper time and the Schwarzschild time only occurs
when the radius is below three times $a_{g}$. In order to
compute the gravitational and Doppler redshifts, we need to separate out the
contribution of each accelerated surface to the radiation spectrum.
This is done by the use of the RORO function. Consider the first RORO function
from $\tau=0$ to $\tau = \tau_{1}$, and the second RORO function from $\tau = \tau_{1}$ to
$\tau = \tau_{2}$. One can show that the complex $\tau_{1}$ poles of these two RORO
functions mutually cancel in their sum,
so the sum of two RORO reduces to one RORO from $\tau = 0$ to $\tau = \tau_{2}$. By breaking
the collapse into discreet radii using the RORO function, one can
fold into the spectrum both the gravitational and Doppler redshift using
the fact that each accelerated radius contributes incoherently to the power
spectrum.

\begin{figure}[htb]
\begin{center}
\leavevmode {\epsfysize=10cm \epsffile{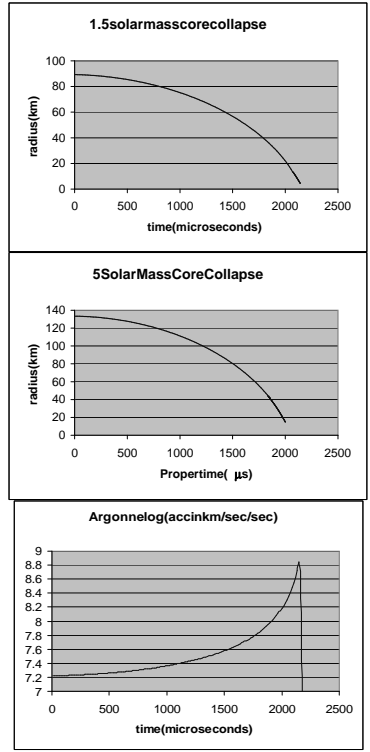}}
\end{center}
\caption[*]{\baselineskip 13pt Radii time trajectory, for core
masses of 1.5 and 5 solar masses, for the initial conditions
stated in the text. Also, an acceleration profile of a core bounce
in the Argonne EOS, for a 0.75 solar mass core.} \label{1471f7}
\end{figure}

In the collapse studies, whenever a star goes below three times $a_{g}$,
we solve for the radius as a function of proper time by means
of the Taylor expansion
  \begin{equation} a(\tau) = a(t) + \frac{da}{dt}(\tau -t) + .5 \frac{d^{2}a}
  {dt^{2}}(\tau -t)^{2}
  \end{equation}
So $a(\tau)$ is reconstructed from three times $a_{g}$ to either
the bounce radius or twice $a_{g}$, for the purposes of calculating the MEMP spectrum.

\begin{figure}[htb]
\begin{center}
\leavevmode {\epsfysize=8cm \epsffile{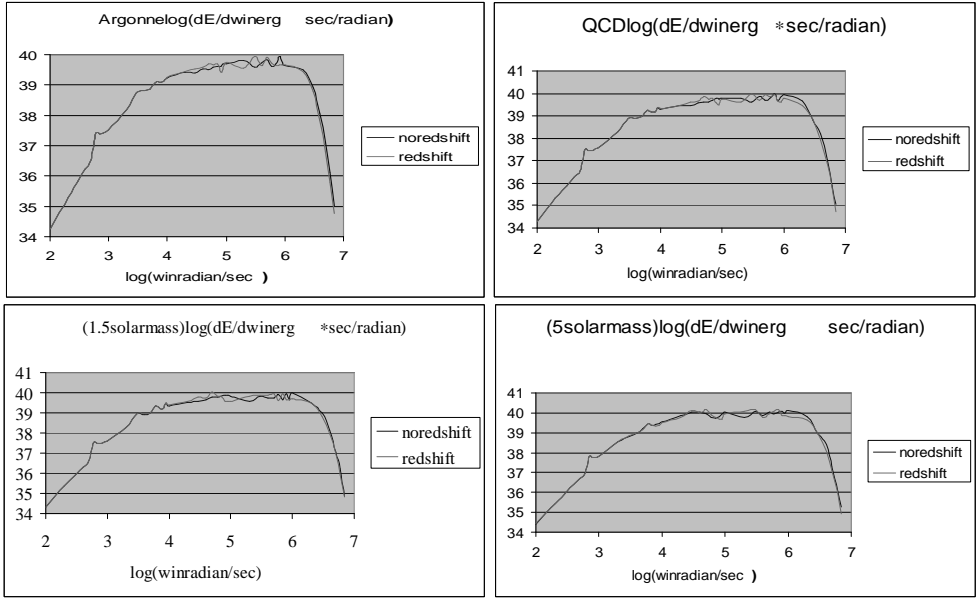}}
\end{center}
\caption[*]{\baselineskip 13pt Spectrum of the radiated energy,
for the Argonne (for a core of 1.0825 solar masses) and QCD (for a
core of 1.3806 solar masses) equations of state, and for a core
mass of 1.5 and 5 solar masses, for the initial conditions
mentioned in the text. } \label{1471f8}
\end{figure}

It is also of
interest to see the kinetics of black hole formation. Unexpectedly, we find
that once the maximum core mass allowing hydrodynamic bounce is exceeded, the
black hole is formed independently of the nuclear EOS. Thus, for example, the
radius function a($\tau$), infall speed $v_{r}(a)$, and acceleration function
acc($\tau$) are all identical for a 1.5 solar mass core collapse,  whatever nuclear
EOS is used. Thus the MEMP of black hole formation is a universal signature. Though
this has been derived in an effective gravitational collapse model, we believe this
is a true theorem in physics. It is of interest then, whether the introduction of stellar
rotation also reintroduces nuclear EOS dependence for black hole MEMP. It is probably true
that once the core mass exceeds the neutron star dynamical mass, whether it is rotating or
not, the black hole MEMP is a universal function. In Fig. 7, we give the kinetics of a 1.5
solar mass core collapse and a 5.0 solar mass core collapse, along with an example of a
core bounce.
\section{Results}

   \begin{table}[h]
   \begin{tabular}{||c|c|c|c|c|c||} \hline
   Core & bounce ? & Initial radius & Final density
   & Final radius  & Time  \\ \hline
   \mbox{} & \mbox{} & \mbox{} & \mbox{} & \mbox{} & \mbox{}  \\
   0.75 & yes  & 70.88 & 6.40(14) & 8.21 & 2175  \\
   1.00 & yes & 78.01 & 7.80(14) & 8.48 & 2145  \\
   1.38 & barely & 86.87 & 7.64(14) & 9.50 & 2110  \\ \hline
   \end{tabular}
   \caption{Results for the QCD equation of state. Core is in $M_{\odot}$,
   Radius in km, time in
   ${\mu}$s, density in $gm/cm^{3}$.
   6.40(14) means 6.40 $\times 10^{14}$.
   The static
   neutron star mass for QCD is 2.34 solar masses.}
   \end{table}

    \begin{table}[h]
    \begin{tabular}{||c|c|c||} \hline
    Core & Emitted MEMP Energy & Received MEMP Energy \\ \hline
    \mbox{} & \mbox{} \mbox{} \\
    QCD 0.75 & 1.07(43) & 1.07(43) \\
    QCD 1.00 & 1.47(43) & 1.47(43) \\
    QCD 1.3806 & 2.05(43) & 1.98(43)\\
    ARG 0.75 & 1.27(43) & 1.27(43) \\
    ARG 1.0825 & 1.82(43) & 1.76(43) \\ \hline
    \end{tabular}
    \caption{The emitted and received (difference due to redshifts)
    directly radiated MEMP energy of the QCD and Argonne (ARG) EOS.
    Units are in ergs.}
    \end{table}

   \begin{table}[h]
   \begin{tabular}{||c|c|c|c|c|c||} \hline
   Core & bounce ? & Initial rad. & Final dens.
   & Final rad. & Time  \\ \hline
   \mbox{} & \mbox{} & \mbox{} & \mbox{} & \mbox{} & \mbox{}  \\
   0.75 & yes  & 70.88 & 1.60(15) & 6.05 & 2195 \\
   1.0825 & barely & 80.10 & 1.29(15) & 7.36 & 2144  \\  \hline
   \end{tabular}
   \caption{Results for the Argonne EOS.
   The static
   neutron star mass for Argonne is 2.10 solar masses.}
   \end{table}

   \begin{table}[h]
   \begin{tabular}{||c|c|c|c||} \hline
   Core & Initial rad.  & Time  & Rad energy  \\ \hline
   \mbox{} & \mbox{} & \mbox{} & \mbox{}  \\
   1.5 & 89.3 & 2142 & 2.66 (43) \\
   5 & 133.40 & 2001 &  7.60(43) \\ \hline
   \end{tabular}
   \caption{1.5 and 5 core mass collapse}
   \end{table}

The code logic is given in Fig. \ref{1471f5}. Following Arnett
(1977), the iron core begins collapse with a fairly uniform
density of $10^{12}$ gm/cm$^{3}$ and initial infall speed $v/c
\sim \frac{2}{3} \times 10^{-2}$. After choosing the mass of the
core, the initial radius is determined.

In Fig. 7, we present the results we obtained for the stellar
surface for core masses of 1.5 and 5.0 solar masses, and in Fig.
\ref{1471f8} the results for the energy spectrum of the emitted
electromagnetic radiation, for the Argonne EOS (for a core of
1.0825 solar masses) and QCD (for a core of 1.3806 solar masses)
EOS cases, and for core masses of 1.5 and 5.0 solar masses
(independent of EOS).

\section{Conclusion}
We have presented an effective gravitational collapse model that
is thought to include the essential physics: spherical radial
infall in the Schwarzschild metric of a homogeneous core of an
advanced star possessing a giant magnetic dipole moment. The
electrodynamic equations are represented by the approximation of
magnetohydrodynamic material response, which will be an excellent
approximation due to the high electrical conductivity. Any nuclear
EOS can be used, and we chose two which have high interest: the
"Argonne" AV14+UVII which represents the typical phenomenological
nuclear EOS and the QCD EOS. Surprisingly, the maximum core mass
that bounced (the dynamic neutron star limit) is rather small
using the phenomenological EOS.

The main result of the paper is the calculation of the MEMP from
the collapse. Both the total energy and its spectral
characteristics are derived. In cases where the core collapse is
greater than the dynamical neutron star mass, the MEMP is the only
electromagnetic experimental signature. It is hoped that this will
spur activity to develop a suitable experimental
receiver\footnote{Due to the peak of the spectral curve around the
wavelength $\sim $ 2km, the receiver will have to be a satellite
able to detect a broadband spectrum within the electromagnetic
pulse. The additional benefit is that terrestrial signals in this
frequency regime cannot penetrate earth's ionosphere, thereby
reducing spurious man-made noise.}. To answer the question as to
whether the surrounding material from the collapsing object will
quench or change an observational signal is clearly complicated,
and it requires a detailed calculation in transport theory using
information on the supernova models that vary tremendously on
their precursor environment, which is beyond the scope of the
present paper.

The existence of two classes of neutron star masses leads to the
situation where a stellar core greater than the dynamical mass,
but less than the static mass, collapses to form a black hole,
thus producing black holes less massive than the mass limit of
neutron star stability. For example, a black hole of 1.6 solar
masses can exist. This possibility sheds light on one of the most
perplexing problems of astrophysics: How do black holes of stellar
masses form? As envisioned here, a core more massive than the
dynamical mass collapses and forms a black hole. It may now
accrete the remaining outlying surface matter to produce a whole
continuum of stellar mass size black holes.

After this paper was submitted for publication, we were informed
of the work of Hanami \cite{Hanami}. This author attempted to
explain gamma-ray bursts using the change in the magnetic field of
a collapsing stellar object. Unfortunately, there are several
problems associated with this work. Hanami did not consider the
fact that the star radiates continuously throughout the collapse
trajectory and his numerical solution violates Maxwell's
equations. He used a zero pressure EOS, which will not give the
correct surface acceleration and core remnant. In order to compute
the total energy, the gravitational redshift must be properly
included. The calculation of the gravitational redshift is a very
difficult problem as explained in the present paper, because, as
the star collapses and radiates, each instantaneous stellar
configuration has a different redshift. In order to do this, one
needs to employ multiple Roll-on-Roll-off mathematical functions
in the complex energy plane, correct for the gravitational
redshift and add the spectrums incoherently. Hanami did not
consider this and simply calculated the redshifted power at the
initial collapse configuration, to obtain his result. At the
bottom of page 688 and top of page 689, Hanami \cite{Hanami}
alludes to the fact that his solution is the quasi-normal ringing
(QNR), but his eigenfrequency is incorrect (QNR have precise
discrete eigenfrequencies). Also his time dependence does not
correspond to any known quasi-normal mode damping.

\end{document}